%% file: MMassi1.tex
\documentclass{evn2002}
\usepackage{graphicx}
\begin{document}
   \title{Investigation of magnetic loop structures in the corona of
 UX\,Arietis}

   \author{M. Massi
          \and
          E. Ros
          }

   \institute{Max-Planck-Institut f\"ur Radioastronomie, Auf dem H\"ugel 69, D-53121 Bonn, Germany
             }

   \abstract{
Most of  the emission coming from the solar corona is confined into closed
magnetic structures in the form of arcs (loops).
Very little is known about the structure of stellar coronae.
The magnetic topology, however, can
be inferred by studying the radio emission coming from electrons trapped in the
magnetic loops.
Evident morphological changes are  produced in fact  by stellar rotation.
We have performed 4 VLBA+Effelsberg runs distributed in time
so as to cover well the
rotational period of 6.44 days of 
the active star UX\,Arietis.
We present here some preliminary results, from those observations.
}

   \maketitle

\section{Large magnetic loops in stellar coronae}

\begin{figure*}[btp]
\includegraphics[width=0.95\textwidth]{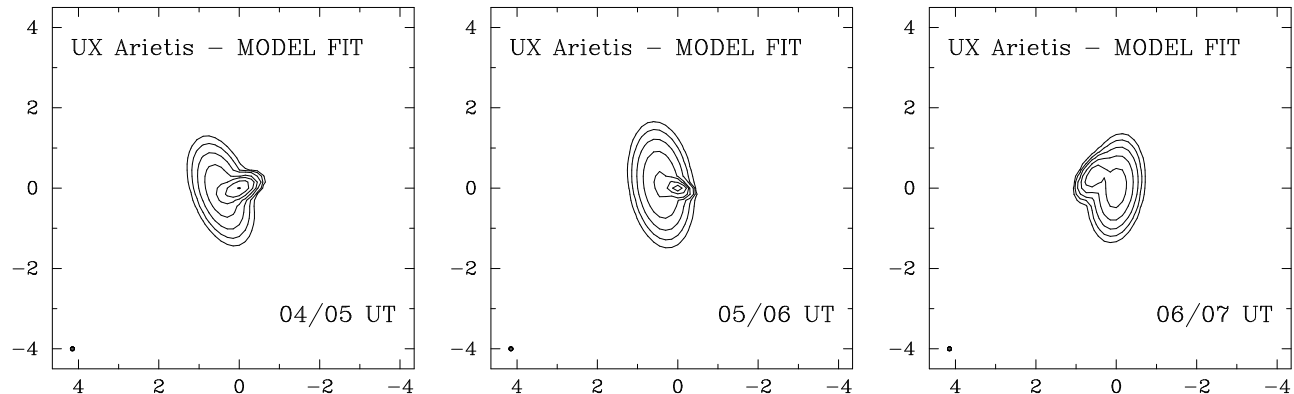}
\caption{ 
Simulated images with the
$(u,v)$-model fitting results of the three  data set of September 26th
corresponding to the UT intervals 04:00-04:50, 
05:00-05:50, and 06:00-06:50, respectively.
 The visibility function of all three data
 sets is well fitted by two Gaussian components.
 The sizes of the two components do not change much from hour
 to hour, but an evident change occur in
 their relative position and orientation.
The origin of this evolution can be explained by 
geometrical factors (i.e.\ star rotation).
}
\end{figure*}

RS\,CVn stars are binary systems 
characterized by intense coronal activity
at X-rays, UV and radio wavelengths. 
One of the most active sources at radio
wavelengths is the system UX\,Arietis
(see Franciosini et al. 1999 and references therein).
Its strong activity is
attributed to magnetic fields which are
  generated in the stellar interior by
a dynamo mechanism,  and emerge from the stellar surface in typical
arc-like structures (loops; see Parker 1955; 1979).
Since the  enhancement of the  magnetic field partially blocks the
 convective energy transport from below, the  foot-points
of an emerging   loop   appear as  dark
spots  in  visible light
(Skinner et al. 1997).
A few and quite large spots on UX\,Arietis have 
been observed  by Vogt \& Hatzes (1991) and Elias et
al.\ (1995): a polar spot and 2 equatorial spots.
The spots are very large in comparison with those on the  Sun,
covering almost 20\% of the stellar surface. Large spots could 
imply  large sizes for the interconnecting  loops.  
Extended loops on RS\,CVn systems, with dimensions much 
larger than the stellar radius are inferred also from
X-ray observations of eclipsing binaries (Walter et al.\ 1988).

Observational  evidence of large structures with sizes comparable to the
binary system comes from  VLBI observations of
UX\,Arietis (Mutel et al.\ 1985; Massi et al.\ 1988;  
Beasley \& Bastian 1996; Massi et al. 1999).
Those studies found  components with sizes of almost 2\,milliarcseconds
(mas),  corresponding to a linear size of 1.5$\times10^{12}$\,cm 
at the distance of 50\,pc), comparable with the orbital size 
of the system (about $1.3\times10^{12}$\,cm). 
The  observed radio emission is
caused by
gyro-synchrotron radiation from mildly relativistic electrons, that 
accelerated during flares  remain trapped inside the magnetic loops.

\section{Magnetic loops anchored on a rotating star}

We have performed VLBA+Effelsberg observations of 
the system UX\,Arietis during four almost 
consecutive days: 23, 25, 26 and 27 September 2001.
We observed at 8.4\,GHz, recording at an aggregate data 
bit rate of 128\,Mb\,s$^{-1}$ with dual polarization.
We interleaved cross-scan measurements at the 100\,m antenna
in Effelsberg during the VLBI observations to 
monitor the total flux density of UX\,Arietis.
We covered a time interval of  six hours for each run, resulting in a 
total $\sim$4\,hr of observing time spent on UX\,Arietis in each run.
The source was flaring during the observations. 
Therefore, in order to distinguish the variability effects 
from structural changes
 we  divide the data into different time slots of 1\,hr length on the basis of
 the Effelsberg  total flux density monitoring.

We have used {\sc difmap} to perform a best-fit of the visibility
data using  elliptical Gaussian components. 
As an example of our procedure we show in Fig.~1 the result of the model fitting
 of  three consecutive data sets
 of September 26th
 corresponding to the UT intervals 04:00-04:50, 
05:00-05:50, and 06:00-06:50, respectively.
A two Gaussian component model is sufficient for all three data sets.
However, as one can see in Fig. 1,  while  the sizes of the two components 
do not change much from hour to hour, 
 their relative positions and orientation evidently  show evident changes.

As discussed above, the optical observations proved the existence of
a large polar spot and of other two spots at the equator.  
Either 
the spots are  rather stable or is stable the longitude/latitude of their 
emergence region is unchanged (Massi et al. 1998).
Therefore, we may interpret the two
fitted Gaussian components as being due to radio emission of electrons trapped
into two magnetic  loops: one "longitudinal" connecting  polar spots,
the other "equatorial" connecting  equatorial spots.
As the star rotates, the loops change their
relative position and orientation with respect to the line of sight, causing
the observed variability of the source structure.
 
This interpretation  can be tested not only by examining
variations from day to day  (as in 
 Franciosini et al.\ 1999 and
in Massi et al.\ 1999) but even in variations from hour to hour.
Already in the 3 maps presented here (of the 16 available maps:
 4\,hours$\times$4\,days), one sees how  
   morphological changes of the source structure
 are consistent with  changes of the relative position of 
the two loops: one of the two 
(the longitudinal one) probably crosses 
the line of sight during the star rotation.
More details will be provided in 
Massi \&  Ros (in preparation). 


\begin{acknowledgements}
We are grateful to the Effelsberg staff for their efforts to guarantee
the success of our observations.
The Very Long Baseline Array of the U.S.\
National Radio Astronomy Observatory is operated by
Associated Universities, Inc., under cooperative agreement with the
U.S.\ National Science Foundation.
\end{acknowledgements}

\end{document}